\documentclass[12pt]{article}
\usepackage{amssymb}
\def\beq{\begin{equation}}
\def\eeq{\end{equation}}
\usepackage[all,2cell,dvips]{xy}

\def\IR{\relax{\rm I\kern -.18em R}}

\begin{document}
\title{Poisson structure and stability analysis of a coupled
system arising from the supersymmetric breaking of Super KdV }
\author{ \Large  A. Restuccia*, A. Sotomayor**}
\maketitle{\centerline{*Departamento de F\'{\i}sica}}
\maketitle{\centerline{Universidad de Antofagasta}
\maketitle{\centerline{*Departamento de F\'{\i}sica}}
\maketitle{\centerline{Universidad Sim\'on Bol\'{\i}var }

\maketitle{\centerline{**Departamento de Matem\'aticas}
\maketitle{\centerline{Universidad de Antofagasta}}
\maketitle{\centerline{e-mail: arestu@usb.ve,
adrian.sotomayor@uantof.cl }}
\begin{abstract}The Poisson structure of a coupled system arising from a supersymmetric
breaking of N=1 Super KdV equations is obtained. The
supersymmetric breaking is implemented by introducing a Clifford
algebra instead of a Grassmann algebra. The Poisson structure
follows from the Dirac brackets obtained by the constraint
analysis of the hamiltonian of the system. The coupled system has
multisolitonic solutions. We show that the one soliton solutions
are Liapunov stable.
\end{abstract}


Pacs: 02.30.Jr, 11.30.Pb, 02.30.Ik, 11.30.-j, 11.10.Ef

\section{Introduction}It is well known that the Korteweg-de Vries
(KdV) equation has two hamiltonians and associated Poisson
structures. The first hamiltonian structure was obtained directly
in the framework of the field $u$ satisfying the KdV equation.
The second hamiltonian structure was obtained on the framework of
the Miura equation (MKdV) and is related to the KdV equation via a
Miura transformation. Although both hamiltonian structures may be
formulated in terms of the $u$ field, the origin of the second
hamiltonian structure relays on the MKdV formulation. Many
extensions of the KdV equation in terms of non-linear coupled
partial differential equations for two or more commutative fields
have been proposed in the literature
\cite{Hirota,Ito,Lou,Yang1,Yang2,Wang}. A particularly
interesting case is the $N=1$ super KdV system \cite{Mathieu1},
which has only one local hamiltonian structure. In this case the
fields take values on the even and odd part of a Grassmann
algebra. When this algebra has finite number of generators, the
supersymmetric system may be formulated directly in terms of the
real coefficients of a field expansion in terms of the basis of
the algebra. One ends up with a coupled system in terms of
commutative real fields. The supersymmetric transformation may
then be defined directly as a transformation on these real fields
which leaves invariant the coupled system.

Recently using such bosonization approach \cite{Adrian} exact
solutions of $N=1$ SKdV equations were obtained \cite{Gao1,Gao2}.

An important aspect of the analysis of the supersymmetric models
is its behaviour under supersymmetric breaking conditions. We
consider here a supersymmetry breaking implemented by replacing
the Grassmann algebra by a Clifford one. This scheme has already
been followed in several works, see for example \cite{Seibert}.

We consider then a coupled system with fields valued on a
Clifford algebra. We find the hamiltonian for such system as well
as its Poisson structure. The hamiltonian formulation on phase
space presents second class constraints. We then introduce Dirac
brackets on the constrained submanifold to define the Poisson
structure for such system. The hamiltonian of this system is
bounded from below which ensures the a physical admissible
content. The solutions of the system satisfy an a priori bound
which suggests the existence of global solutions to the system.

Although the system has multisolitonic solutions it does not have
an infinite number of local conserved quantities.

We finally consider the stability analysis of the one-soliton
solutions of the extended KdV system arising from $N=1$ SKdV
under supersymmetric breaking conditions. Following the approach
introduced by \cite{Benjamin} and \cite{Bona} we show that the
one-soliton solutions are Liapunov stable, for details see
\cite{Adrian2}.

\section{Breaking supersymmetry on SKdV} The $N=1$ SKdV equations
have an infinite sequence of local conserved quantities
\cite{Mathieu1} as well as an infinite sequence of non-local
conserved quantities \cite{Dargis,Andrea1,Andrea2}. To break
supersymmetry we consider fields $u$ and $\xi$ valued on a
Clifford algebra instead of being Grassmann algebra valued. We
thus take $u$ to be a real valued field while $\xi$ to be an
expansion in terms of the generators $e_i,i=1,\ldots$ of the
Clifford algebra: \beq\xi=\sum_i
\varphi_ie_i+\sum_{ij}\varphi_{ij}e_ie_j+\sum_{ijk}\varphi_{ijk}e_ie_je_k+\cdots\label{eq1}\eeq
where \beq e_ie_j+e_je_i=-2\delta_{ij}\label{eq2}\eeq and
$\varphi_i,\varphi_{ij},\varphi_{ijk},\ldots$ are real valued
functions. We define $ \bar{\xi}=\sum_{i=1}^\infty
\varphi_i\bar{e_i}+\sum_{ij}\varphi_{ij}\bar{e_j}\bar{e_i}+\sum_{ijk}\varphi_{ijk}\bar{e_k}\bar{e_j}\bar{e_i}+\cdots$
where $\bar{e_i}=-e_i$. We denote as in superfield notation the
body of the expansion those terms associated with the identity
generator and the soul the remaining ones. Consequently the body
of $ \xi\bar{\xi}$, denoted by $ \mathcal{P}(\xi\bar{\xi})$, is
equal to
$\Sigma_i\varphi_i^2+\Sigma_{ij}\varphi_{ij}^2+\Sigma_{ijk}\varphi_{ijk}^2+\cdots$
In what follows, without loss of generality, we rewrite
$\mathcal{P}(\xi\bar{\xi})=\Sigma_i
\varphi_i^2+\Sigma_{ij}\varphi_{ij}^2+\Sigma_{ijk}\varphi_{ijk}^2+\cdots$
simply as $\mathcal{P}(\xi\bar{\xi})=\Sigma_i\varphi_i^2$.

The resulting system after the breaking of supersymmetry becomes
\beq\begin{array}{l}u_t=-u^{\prime\prime\prime}-uu^\prime-\frac{1}{4}{\left(\mathcal{P}\left(\xi\bar{\xi}\right)\right)}^\prime
\\ \xi_t=-\xi^{\prime\prime\prime}-\frac{1}{2}{(\xi u)}^\prime.
\end{array}\label{eq3}\eeq
This coupled KdV system arises from the following Lagrangian,
expressed in terms of the fields $w$ and $\eta_1$ related to $u$
and $\varphi_i$ by \beq u=w^\prime \mathrm{\:and\:}
\varphi_i=\eta_i^\prime,\label{eq4}\eeq \beq S(w,\eta_i)\equiv
\int dxdt\left[\frac{1}{2}w^\prime
\partial_tw+\frac{1}{6}{\left(w^\prime\right)}^3-\frac{1}{2}{\left(w^{\prime\prime}\right)}^2+\frac{1}{4} w^\prime
{\left(\eta_i^\prime\right)}^2+\frac{1}{2}\eta_i^\prime\partial_t\eta_i-\frac{1}{2}{\left(\eta_i^{\prime\prime}\right)}^2
\right],\label{eq5}\eeq where a repeated index $i$ implies
summation on that index. Variations of the action $S(w,\eta_i)$
yields the corresponding field equations (3).

The hamiltonian and corresponding Poisson structure associated to
the action (5) follows by introducing the conjugate momenta
associated to $w,\eta_i$. We will denote them by $p,\sigma_i$
respectively \beq \begin{array}{lll}p:=\frac{\partial
\mathcal{L}}{\partial
(\partial_tw)}=\frac{1}{2}w^\prime=\frac{1}{2}u \\ \\
\sigma_i:=\frac{\partial \mathcal{L}}{\partial (\partial_t
\eta_i)}=\frac{1}{2}\eta_i^\prime=\frac{1}{2}\varphi_i.
\end{array} \label{eq6}\eeq (6) are primary constraints on the
phase space. The hamiltonian may be obtained by performing a
Legendre transformation.

We obtain \beq H=\int_{-\infty}^{+\infty}
\left(-\frac{1}{3}u^3-\frac{1}{2}u\mathcal{P}(\xi\bar{\xi})+{(u^\prime)}^2+\mathcal{P}(\xi^\prime\bar{\xi^\prime})\right)dx.
\label{eq7}\eeq Following the Dirac approach for constrained
systems it turns out that (6) are the only constraints on the
phase space, and they are second class constraints. The Poisson
structure on the constrained submanifold of phase space is then
given by the Dirac brackets. They are given by \beq\begin{array}{lll}\left\{u(x),u(y)\right\}_{DB}=\partial_x\delta(x,y),\\
\left\{\varphi_i(x),\varphi_j(y)\right\}_{DB}=\delta_{ij}\partial_x\delta(x,y),\\\left\{u(x),\varphi_i(y)\right\}_
{DB}=0.\end{array}\label{eq8}\eeq We thus have obtained the
hamiltonian $H$ and the Poisson structure of the coupled KdV
system (3). Besides the hamiltonian $H$ there are other conserved
quantities under the evolution determined by the coupled KdV
system (3). The following are conserved quantities
\beq\begin{array}{llll}\hat{H_{\frac{1}{2}}}=\int_{-\infty}^{+\infty}\xi
dx,\\ \hat{H_1}=\int_{-\infty}^{+\infty} u dx,
\\ V\equiv
\hat{H_3}=\int_{-\infty}^{+\infty}\left(u^2+\mathcal{P}(\xi\bar{\xi})\right)dx.
\end{array}\label{eq9}\eeq It is interesting to notice that the
non-local quantity \beq\int_{-\infty}^\infty
\xi(x)\int_{-\infty}^x\xi(s)dsdx\label{eq10}\eeq is also
conserved. Remarkably, this quantity in terms of a Grassmann
valued $\xi$ is also conserved for the $N=1$ SKdV system.

The system (3) besides being not invariant under supersymmetry,
by construction, it does not have an infinite sequence of local
conserved quantities. It does have, however, multisolitonic
solutions. We discuss now the stability properties of the
one-soliton solutions.

\section{Stability of the solitonic solutions} An important first
step in the analysis is to show an a priori bound \cite{Tao} for
the solutions of the system (3).

We denote by $\|\hspace{2mm} \|_{H_1}$ the Sobolev norm \beq
{\|(u,\xi)
\|_{H_1}}^2=\int_{-\infty}^{+\infty}\left[\left(u^2+\Sigma_{i=0}^\infty
\varphi_i^2 \right)+\left ({u^\prime}^2+\Sigma_{i=0}^\infty
{\varphi_i^\prime}^2\right)\right]dx. \label{eq11} \eeq We obtain
from the expressions of $V$ and $H$
 \beq {\|(u,\xi)
\|_{H_1}}^2\leq
V+H+\frac{1}{2}\int_{-\infty}^{+\infty}|u|\left(u^2+\mathcal{P}\left(\xi\bar{\xi}\right)\right)dx.\label{eq12}
\eeq We may now use \beq \sup|u|\leq \frac{1}{\sqrt{2}}{\|u
\|_{H_1}}\leq \frac{1}{\sqrt{2}}{\|(u,\xi)
\|_{H_1}}\label{eq13}\eeq to get \beq {\|(u,\xi) \|_{H_1}}^2\leq
V+H+\frac{1}{2\sqrt{2}}V{\|(u,\xi) \|_{H_1}}.\label{eq14} \eeq It
then follows \beq{\|(u,\xi) \|_{H_1}}\leq
\frac{d+\sqrt{d^2+4e}}{2}\label{eq15}\eeq where
$d=\frac{1}{2\sqrt{2}}V$ and $e=V+M.$ We notice that
$d^2+4e\geq0.$

Given $V$ and $H$ from the initial data and a solution satisfying
those initial conditions, then ${\|(u,\xi) \|_{H_1}} $ is bounded
by (15) for all $t\geq0$. This is an a priori bound and a strong
evidence of the existence of the solution for $t\geq0$.

We consider the stability of a solution $( \hat{u},\hat{\xi})$ of
(3) in the Liapunov sense:

$( \hat{u},\hat{\xi})$ is stable if given $\epsilon$ there exists
$\delta$ such that for any solution $(u,\xi)$ of (3), satisfying
at $t=0$

\beq
d_I\left[\left(u,\xi\right),\left(\hat{u},\hat{\xi}\right)\right]<\delta\label{eq16}
\eeq then \beq
d_{II}\left[\left(u,\xi\right),\left(\hat{u},\hat{\xi}\right)\right]<\epsilon\label{eq17}\eeq
for all $t\geq 0$.

We consider first the stability of the ground state solution $
\hat{u}=0,\hat{\xi}=0.$ We take $d_I$ and $d_{II}$ to be the
Sobolev norm $\|(u-\hat{u},\xi-\hat{\xi}) \|_{H_1}.$

We get \[V\leq \|(u,\xi) \|^2_{H_1}\] and
\[H\leq \int_{-\infty}^{+\infty}\left(\frac{1}{2}|u|\left(u^2+\mathcal{P}\left(\xi\bar{\xi}\right)\right)+{u^\prime}^2
+\mathcal{P}\left(\xi^\prime\bar{\xi}^\prime\right)\right)dx \leq
\frac{1}{2\sqrt{2}}{\|(u,\xi) \|_{H_1}}^3+{\|(u,\xi)
\|_{H_1}}^2.\] It then follows from the a priori bound (15) the
stability of the ground state $ \hat{u}=0,\hat{\xi}=0.$

We now consider the stability of the one-soliton solution $
\hat{u}=\phi,\hat{\xi}=0,$ where $\phi$ is the one-soliton
solution of KdV which satisfies \beq
\phi^{\prime\prime}+\frac{1}{2}\phi^2=\mathcal{C}\phi.
\label{eq18} \eeq

 The proof of stability is based on
estimates for the $u$ field which are analogous to the one
presented in \cite{Benjamin,Bona} while a new argument will be
given for the $\xi$ field. The distances we will use are

\beq
d_I\left[\left(u_1,\xi_1\right),\left(u_2,\xi_2\right)\right]=\|\left(u_1-u_2,\xi_1-\xi_2\right)\|_{H_1}
\eeq \beq
d_{II}\left[\left(u_1,\xi_1\right),\left(u_2,\xi_2\right)\right]=\inf_\tau\|\left(\tau
u_1-u_2,\xi_1-\xi_2\right)\|_{H_1} \eeq where $\tau u_1$ denotes
the group of translations along the $x$-axis. $d_{II}$ is a
distance on a metric space obtained by identifying the
translations of each $u\in H_1( \mathbb{R})$ \cite{Benjamin}.
$d_{II}$ is related to a stability in the sense that a solution
$u$ remains close to $ \hat{u}=\phi$ only in shape but not
necessarily in position.

At $t=0$ we assume that \beq
d_I\left[\left(u,\xi\right),\left(\phi,0\right)\right]=\|\left(h,\xi\right)\|_{H_1}<\delta.
\label{eq18} \eeq Using estimates as in \cite{Benjamin,Bona} one
can show that \beq|\Delta
H|\leq\left[\max\left(1,\mathcal{C}\right)+\frac{1}{3\sqrt{2}}\delta\right]{\|\left(h,\xi\right)\|}_{H_1}^2\label{eq19}\eeq
where $h\equiv u-\phi.$

The next argument introduces besides the ideas in
\cite{Benjamin,Bona}, new estimates on $\xi$, see \cite{Adrian2}.
After several delicate estimates we obtain \beq\Delta
H\geq\frac{1}{6}l{\left\{ d_{II}\left[(u,\xi),(\phi,0)\right]
\right\}}^2\label{eq20} \eeq where $l=\min(1,\mathcal{C})$.

The upper and lower bounds for $\Delta H$ imply the stability of
the one-soliton solution of the coupled KdV system.

\section{Conclusions} By breaking the supersymmetry in the $N=1$
SKdV equations we arrive to a coupled KdV system. We found its
hamiltonian and associated Poisson structure. The hamiltonian is
bounded from below and consequently has an admissible physical
interpretation. This important property is then used to show the
stability of the ground state solution as well as the one-soliton
solutions of the coupled KdV system. We determine also an a
priori bound giving a strong evidence of the existence of
solutions for all $t\geq0.$

The property of having a hamiltonian bounded from below is very
important and it is shared with the hamiltonian of the super KdV
equations. However, the stability of the solitonic solutions of
$N=1$ super KdV equations has, so far, not been proven in the
literature.

 $\bigskip$

\textbf{Acknowledgments}

A. S. and A. R. are partially supported by Project Fondecyt
1121103, Chile.

\end{document}